# Reconfigurable On-chip Photoconductive Switches

Cheng-Yi Fang,[1] Hung-Hsi Lin,[1] Mehdi Alouini,[3] Yeshaiahu Fainman,[2] and Abdelkrim El Amili[2,*]

1. University of California San Diego, Materials Science & Engineering Program, La Jolla, CA 92093
2. University of California San Diego, Department of Electrical & Computer Engineering, La Jolla, CA 92093
3. Institut FOTON, University of Rennes 1, CNRS, Campus de Beaulieu, Rennes, France

Corresponding Author: Dr. Abdelkrim El Amili, aelamili@ucsd.edu

## Abstract

Microwave photonics uses light to carry and process microwave signals over a photonic link. However, light can instead be used as a stimulus to microwave devices that directly control microwave signals. Such optically controlled amplitude and phase-shift switches are investigated for use in reconfigurable microwave systems, but they suffer from large footprint, high optical power level required for switching, lack of scalability and complex integration requirements, restricting their implementation in practical microwave systems. Here, we report Monolithic Optically Reconfigurable Integrated Microwave Switches (MORIMSs) built on a CMOS compatible silicon photonic chip that addresses all of the stringent requirements. Our scalable micrometer-scale switches provide higher switching efficiency and require optical power orders of magnitude lower than the state-of-the-art. Also, it opens a new research direction on silicon photonic platforms integrating microwave circuitry. This work has important implications in reconfigurable microwave and millimeter wave devices for future communication networks.

# 1. Introduction

Reconfigurability has become a crucial feature in modern, agile, microwave and millimeter wave (MMW) systems for emerging wireless communications, sensing and imaging.[1–5] Among various existing building blocks, optically reconfigurable MMW amplitude and phase-shift switches are key devises for beam steering in RADAR systems and reconfigurable antennas for emerging 5G wireless communications network [6,7]. An optically controlled switch is a device whose electrical state can be tuned from insulating (Off state) to conductive (On state) by means of optical stimuli[8–11]. The underlying physics relies on photoconductive effect that occurs through the light interaction with a semiconductor material[12]. The illumination with a photon energy larger than the semiconductor bandgap generates electron-hole pairs in the control layer which modifies its electrical conductivity and affects the amplitude and phase of MMW signals.

The idea of using light to control or even introduce signals directly into microwaves devices[7,13,14] has drawn great interest in the microwave community driven by the need for dynamic control, fast response, immunity to electromagnetic interference, and good isolation between the controlling and controlled devices. The optical solution promises better performances compared to its classical analogue that utilizes electrical or microelectromechanical system which are prone to signal distortion and unwanted electromagnetic interferences.[1] Various reconfigurable microwave functionalities have been demonstrated including cognitive radio applications[15], microwave mixers [16] and phase shifters [17,18]. Although optically controlled microwave amplitude and phase switches have attracted appreciable attention due to their superior potential performances, they are not yet sufficiently advanced for implementation in practical microwave systems. The main reasons are twofold: (i) lack of scalability and compactness due to the fact that current approaches use free-space or fiber illumination[19,20] thus requiring costly and complex packaging and (ii) the optical power level required to perform a switching operation [9,10,21–23] is prohibitively high, e.g., to achieve On/Off

RF switching with extinction ratio of ~10dB requires optical power in the range of tens to several hundreds of a milliwatts. Moreover, it should be noted that photodiode and phototransistors switches can operate at low optical power but they require electrical bias and are not scalable in large high-frequency phased array systems.[24,25] These challenges can be addressed by utilizing photonic technology to manipulating MMW signals in microwave systems.

In this manuscript we overcome these challenges and report the design, fabrication and experimental demonstration of Monolithic Optically Reconfigurable Integrated Microwave Switches (MORIMSs) built on a CMOS compatible silicon photonic chip. Silicon nitride waveguides are exploited to route optical waves towards silicon photoconductive patches to switch microwave signals at different locations of the chip. Photonic integration allows high light coupling efficiency into silicon photoconductive patches. We show that the integration of microwave circuits and optical waveguides on a CMOS platform provides scalable micrometer-scale footprint switches with higher switching efficiency, large phase shift and optical power requirement orders of magnitude lower than the state of the art. Our work paves the way for a new generation of complex optically reconfigurable microwave circuits that benefit from the integrated silicon photonics technology.

## 2. MORIMS architectures

Emerging photonic integrated circuits (PICs) technology [26] has already made a significant impact on high-speed optical interconnects and digital optical communication links[27]. PICs manufacturing using silicon on insulator (SOI) platform is compatible with CMOS process allowing mass production at low cost[28]. It offers highly desirable features such as small footprint, scalability and reduced power consumption. By taking advantages of integrated photonics flexibility, our proposed devices use one single waveguide to control multiple microwave switches in different locations on the chip. Moreover, integrated photonics offers the possibility to engineer and optimize light coupling efficiency from optical waveguide to silicon photoconductive patches in order to achieve high switching performance. Depending on the application, the microwave switches can also be addressed independently or combined with a variety of photonic building blocks such as Y-branch, directional couplers, ring resonators, Mach-Zehnder modulators, etc. With this vision in mind, we have developed two different MORIMS architectures as illustrated in Fig. 1a and b to meet different demands. Both architectures use a single mode silicon nitride ($SiN_x$) waveguide, silicon (Si) photoconductive patch and aluminum (Al) co-planar waveguide transmission lines all built on the same SOI wafer. The signal electrode gap is made of a Si photoconductive patch that acts as an electrical insulator (Off state) but under illumination acts as a conductor (On state). The MORIMS operates with optical radiation at the wavelength of 808 nm.

The SOI wafer consists of 250nm-thick device layer and 3μm-thick buried oxide layer. During the fabrication process, most of the silicon material is removed to form Si photoconductive patches with dimension of 16μm by 12μm. Single-mode $SiN_x$ ridge waveguide with the dimensions of 800nm-width and 400nm-height are fabricated and used to guide light toward Si patches in order to activate them at different locations on the chip. The ridge waveguide and Si photoconductive patch are cladded by 1μm-thick $SiO_2$ layer. The Ground-Signal-Ground (GSG) transmission lines consist of 800nm-thick Al lines with a tapered signal electrode toward the Si photoconductive patch.

The two proposed structures, referred as "tapered" and "through" type, correspond to the way the optical waveguide is designed on top of the silicon photoconductive patch to optically control its conductivity. The "tapered type" structure (Fig. 1a), where $SiN_x$ waveguide is tapered on the Si photoconductive patch, is devoted to maximizing the coupling of light from $SiN_x$ waveguide to Si photoconductive patch. The tapered-type structure allows ~84% of the energy to be coupled into the Si photoconductive patch. The "through type (Fig. 1b), where waveguide crossing the Si photoconductive patch, can be utilized in cascaded configuration, i.e., connecting "optically" different microwave circuits as it will be demonstrated later. Indeed, this configuration allows ~67% of the energy to be coupled into the silicon patch while the remaining light can be used to control the following microwave circuit.

Fig. 1c and d show the SEM images of MORIMSs of both types. The $SiN_x$ waveguide conformally covers the Si photoconductive patch without any crack and discontinuity.

## 3. Performance of MORIMSs

The On/Off performances of the MORIMS are characterized by measurements of the S-parameters. Fig. 2a and b show the measured $S_{21}$ parameter of tapered- and through-type structures at On and Off state up to ~40GHz. To characterize the switches performance, the extinction ratio $R_{on/off}=|S_{21}(On)/S_{21}(Off)|$ is adopted as the figure of merit that qualifies amplitude switching efficiency for a given microwave frequency[10]. Fig. 2c and d show $R_{on/off}$ with respect of input optical power at frequencies of 5GHz, 20GHz and 40GHz. Overall, the On/Off ratios increase linearly from 0 to ~1.5mW before reaching a saturation plateau. As expected, the tapered-type switch shows higher performance, with switching efficiency of ~25dB and ~23dB at 5GHz and 20GHz, respectively compared to ~14dB and ~12 dB achieved at same frequencies with the through-type configuration. Although the through-type is less efficient under same incident optical power, the remaining energy in the waveguide can be used to control another switch as shown next. It is worth mentioning that the switching time of the proposed device is on the order of few micro-seconds which is compatible with beam steering and beamforming applications requirements.

Table 1 shows the state-of-art photoconductive switches in terms of switching performances, optical power requirement and footprint. Since most of the literature has reported switching at low frequencies and few demonstrations have been done at very high frequencies, the amplitude switching performances are thus compared at frequencies below and above 10GHz. Remarkably, MORIMSs provide higher performances, i.e., ~29dB ~25dB, ~ 23dB and 11dB switching efficiency at 1, 5, 20 and 40GHz respectively, while using less than 2mW which is by orders of magnitude lower than free-space illumination-based switches. Moreover, MORIMS shows the capability of on-chip integration which can be incorporated into complex on-chip photonics and microwave system with ultra-compact footprint to meet the desired high- packing density.

## 4. Performances of cascaded MORIMSs

To demonstrate scalability and integration of multiple reconfigurable switches on the same chip, three MORIMSs were designed and fabricated as depicted in Fig. 3a. The MORIMSs in series and parallel configuration are fed by one single input optical waveguide. The injected light is routed toward two different paths using a 3-dB Y-branch coupler. One of the paths addresses two cascaded through-type MORIMSs. Fig. 3b - d show $R_{on/off}$ at different locations. Because MORIMS_1 and 3 are in parallel, they show same performance, for instance, their switching efficiency reaches ~10dB at 20GHz. However, the switching efficiency of MORIMS_2 in series with MORIMS_1, drops by only ~ 4 dB at 20GHz. MORIMS shows promising performances for cascaded optically reconfigurable switches for frequency and phased array system.

## 5. Discussion and summary

The proposed optically reconfigurable switches are a proof of concept that can be easily implemented in beamforming and beam steering microwave systems which require moderate switching time constant. Moreover, the proposed integrated devices could also enable more advance functionalities when combining other well-established photonic building blocks such as ring resonators, directional couplers and Mach-Zehnder modulators on the same chip. The proposed approach can be tailored in the future generation of ultra-high frequency communications systems which will face stringent requirements in terms of frequency bandwidth, power consumption, size and packing density, and low-cost for mass production. In that area, ultra-fast photoconductive switches exploiting III-V materials, with ultra-short carrier lifetime, are required and outstanding efforts has been already made[8,29]. The proposed approach could be exploited in sampling application that requires the combination of several switches with very accurate time delays between them. This work is a real added value for developing integration technology for microwave signal processing.

In summary, we have demonstrated monolithic optically reconfigurable integrated microwave switches on a SOI chip. Our approach consists of co-integration of microwave circuits with integrated photonic devices to form optically reconfigurable microwave switches. A single input $SiN_x$ waveguide is used to route the light toward switches at different location on chip. Integrated photonics provides miniaturized Si photoconductive patches, high confinement of light in the waveguide and high coupling efficiency of light from waveguide to silicon photoconductive microwave switches. Consequently, the demonstrated engineered devices outperform their classical analogues in term of On/Off switching efficiency, footprint and optical power level requirement. We experimentally demonstrate high microwave amplitude switching performances of over 25dB around 5GHz, 23dB around 20GHz and 11dB at 40GHz, and lower optical power requirement (~ 2mW) by orders of magnitude lower than the state-of-art photoconductive switches. Scalability is a challenge that has been also advanced by demonstrating integrated multiple reconfigurable switches on the same SOI chip with high amplitude switching performance. Moreover, phase shifts of 20° and 60° were measured for microwave signals at 20GHz and 40GHz, respectively. This work is an important step in introducing photonics into direct processing of microwave signals, paving the way towards optically reconfigurable microwave and millimeter wave devices for future ground, embedded radar systems, and emerging 5G wireless communication networks.

**Acknowledgments:** This work was supported by the Defense Advanced Research Projects Agency (DARPA) and DARPA NLM, the Office of Naval Research (ONR) Multidisciplinary University Research Initiative (MURI), the National Science Foundation (NSF) Grants DMR-

1707641, CBET-1704085, ECCS-1405234, ECCS-1644647, CCF-1640227 and ECCS-1507146, the NSF ERC CIAN, the Semiconductor Research Corporation (SRC), the NSF's NNCI San Diego Nanotechnology Infrastructure (SDNI), Chip-Scale Photonics Testing Facility (CSPTF), Nano3, the Army Research Office (ARO), and the Cymer Corporation.

The authors acknowledge Dr. N. Rostomyan and Dr. A. Smolyaninov for the fruitful discussions.

## (a) Tapered Type

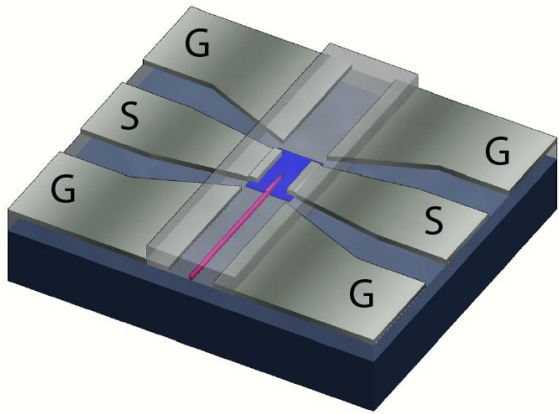

## (b) Through Type

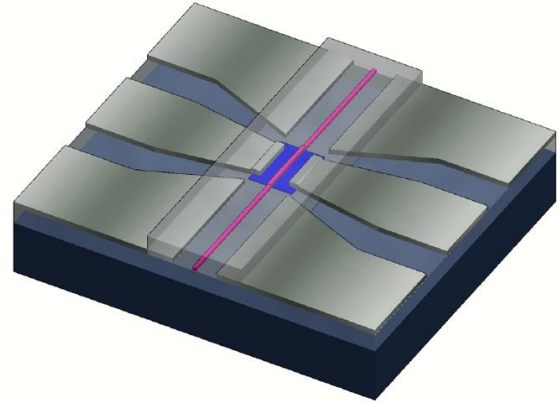

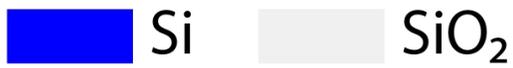
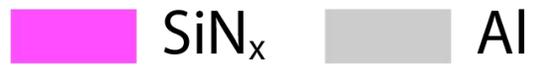

## (c)

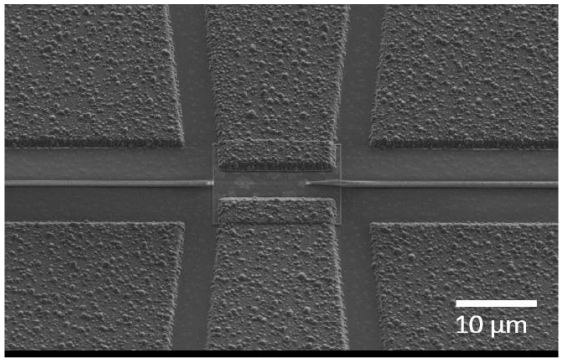

## (d)

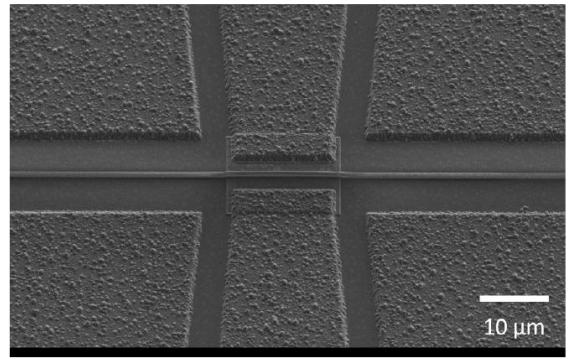

Fig. 1. Schematic and SEM images of Monolithic Optically Reconfigurable Integrated Microwave Switches (MORIMSs). (a) Tapered type: $SiN_x$ waveguide tapered on top of photoconductive Si patch; (b) Through type: $SiN_x$ waveguide not tapered and connected to the output port. (c) side view SEM image of tapered type structure shown in Fig. 1a and (d) side view SEM image of through type structure shown in Fig. 1b.

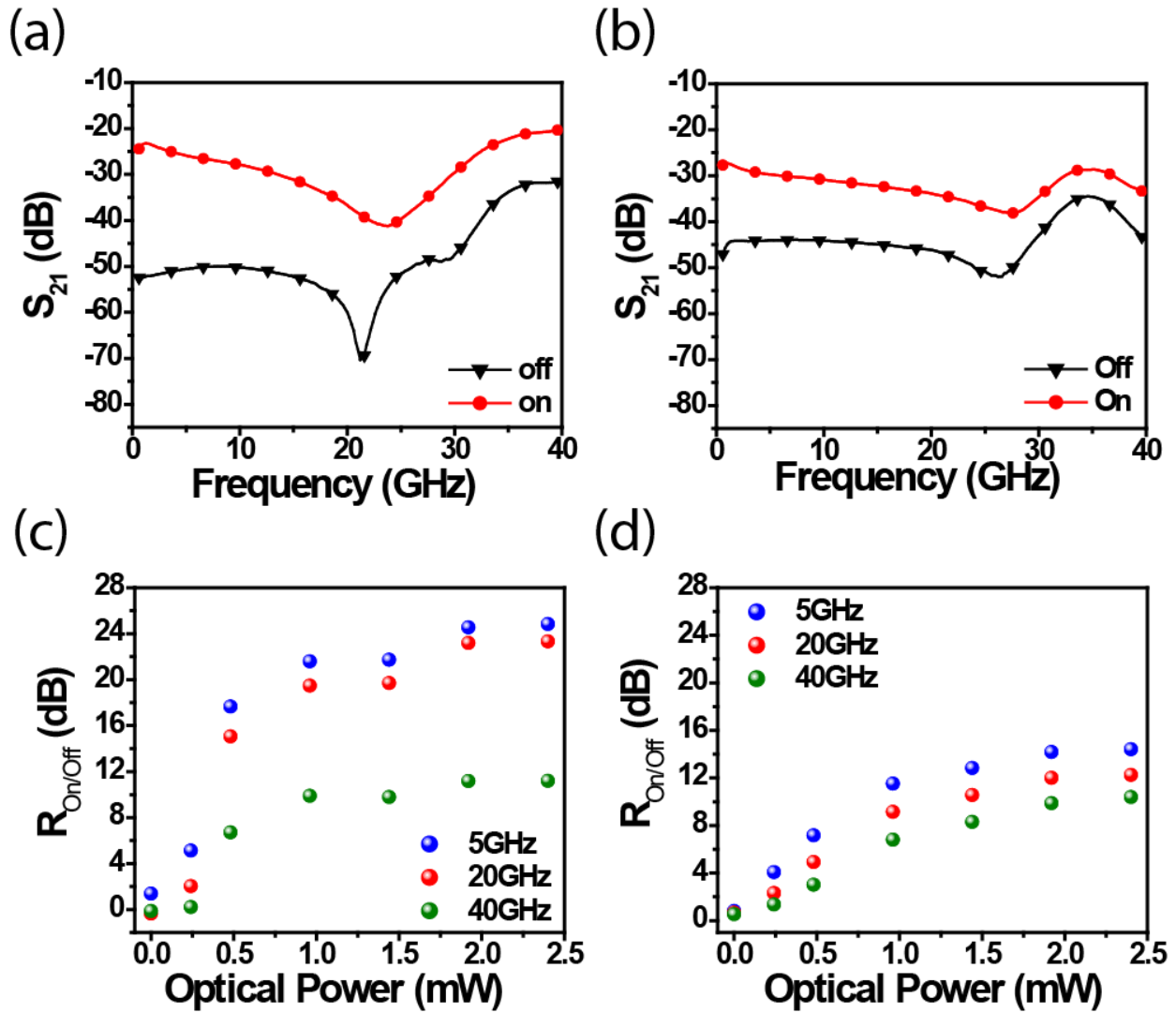

Fig. 2. Measured S$_{21}$ of MORIMS of (a) tapered type (b) through type; R$_{on/off}$ with respect to incident optical power of (c) tapered type (d) through type.

(a)

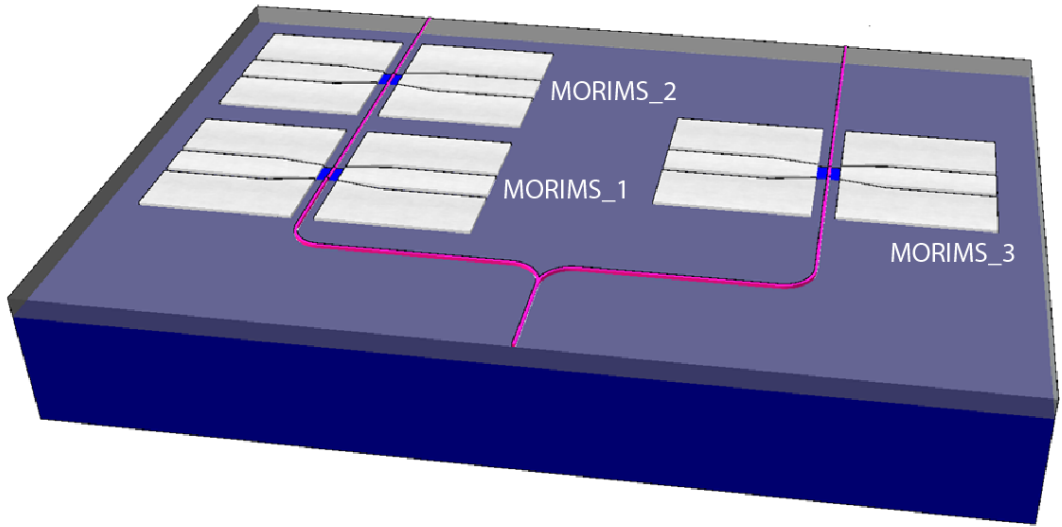

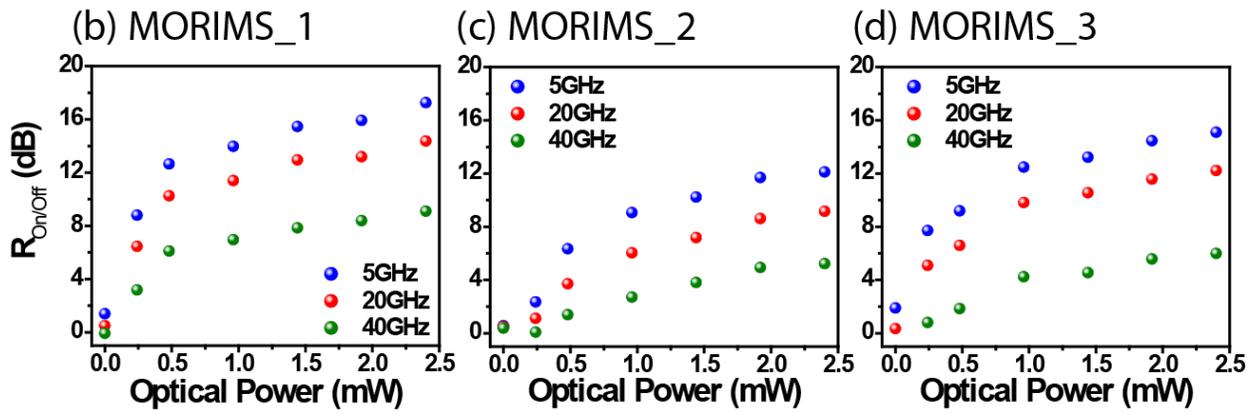

Fig. 3. (a) Schematic of a basic MORIMS circuit. The $R_{on/off}$ at 5, 20 and 40GHz with respect to incident optical poweer of (b) MORIMS_1, (c) MORIMS_2 and (b) MORIMS_3.

Table 1 Different microwave photoconductive switches with their reported frequency, S-parameter on/off ratio, power consumption and device footprint.

| Year [Ref] | $R_{on/off}$ (dB) (f≤10GHz) | $R_{on/off}$ (dB) (f>10GHz) | Optical Power requirement (mW) | Footprint | Photoconductive material | On-chip integration |
|---|---|---|---|---|---|---|
| 1995[30] | 45 (1.7GHz) | | 143 | 10 μm×1.6cm | GaAs | No |
| 2003[31] | | 15.4 (20GHz) 8.7 (35GHz) | 90 | Gap: 130 μm | GaAs | No |
| 2003[32] | 25 (1GHz) | | 15 | 1.2mm×1.4mm× 0.6mm | GaAs | No |
| 2006[22] | 15 (2GHz) | | 200 | 1mm×2mm× 0.3mm | Si | No |
| 2006[33] | | 2.9 (40GHz) | 100 | Not report | GaAs | No |
| 2009[9] | 27.4 (2GHz) | | 40 | $0.25cm^2$×0.5cm | Si | No |
| 2010[8] | 9 (1.5GHz) | | 80 | 100μm×5μm | GaNAsSb | No |
| 2012[34] | 18 (3GHz) | | 200 | 1mm×2mm× 0.28mm | Si | No |
| 2012[10] | | | 100 | 0.1μm×0.1μm× 150μm | GaAs | No |
| 2015[23] | 9 (3.5GHz) | | 20 | 3mm×2mm× 0.28mm | Si | No |
| 2016[35] | 5 (10GHz) | | 62 | ~1 μm×1 μm | Black Phosphorous | No |
| This work | 29 (1GHz) 25 (5GHz) | 23 (20GHz) 11 (40GHz) | 2 | 12μm×16μm× 250nm | Si | Yes |